# Tailoring the ferromagnetic surface potential landscape by a templating two-dimensional metal-organic porous network

Lu Lyu[†,*], Martin Anstett[†], Ka Man Yu[†], Azadeh Kadkhodazadeh[†], Martin Aeschlimann[†], Benjamin Stadtmüller[†,§,*]

[1]Department of Physics and Research Center OPTIMAS, Rheinland-Pfälzische Technische Universität Kaiserslautern-Landau, Erwin-Schrödinger-Straße 46, 67663 Kaiserslautern, Germany.

[2]Institute of Physics, Johannes Gutenberg University Mainz, Staudingerweg 7, 55128 Mainz, Germany.

[*]Email: llyu@rhrk.uni-kl.de
[*]Email: bstadtmueller@physik.uni-kl.de

**Abstract**

Two-dimensional metal-organic porous networks (2D-MOPNs) have been identified as versatile nanoarchitectures to tailor surface electronic and magnetic properties on noble metals. In this context, we propose a protocol to redecorate a ferromagnetic surface potential landscape using a 2D-MOPN. Ultrathin cobalt (Co) films grown on Au(111) exhibit a well-ordered surface triangular reconstruction. On the ferromagnetic surface, the adsorbed 2,4,6-tris(4-pyridyl)-1,3,5triazine (T4PT) molecules can coordinate with the native Co atoms to form a large-scale Co-T4PT porous network. The Co-T4PT network with periodic nanocavities serves as a templating layer to reshape the ferromagnetic surface potential. The subsequently deposited $C_{60}$ molecules are steered by the network porous potential and the neighboring $C_{60}$ interactions. The prototype of the ferromagnetic-supported 2D-MOPN is a promising template for the tailoring of molecular electronic and spin properties.

**Introduction**

Two-dimensional metal-organic porous networks (2D-MOPNs) have been sought-after nanoarchitectures for exotic tessellation patterns[1], catalytic reactions[2], magnetic exchange[3,4] and surface quantum confinements[5]. The protocol for fabricating a low-dimensional metal-organic architecture is based on the surface self-assembly between organic molecules and metal adatoms[6]. The selected molecule with an electronegative ligand in the end-group activates a robust coordination with the surface metal adatoms. Such a directional metal-ligand interaction can trigger the formation of a long-range ordered porous nanoarchitecture. To enhance surface self-assembly, 2D-MOPNs are typically fabricated on weakly interacting substrates such as low-index noble metals[6] and 2D materials[7]. Each organic molecule on the substrates can form an effective potential barrier that blocks the penetration of surface electrons and atoms[5], analogously to the adatom-induced scattering[8]. Therefore, the periodic nanocavities in a 2D-MOPN are a quantum well array, in which the porous potential can capture the electrons[9], adatoms[10] and even molecules[11]. The self-assembly of a 2D-MOPN alters the surface potential landscape of the substrate, thus dictating the arrangement of subsequent adsorbates.

The organic/ferromagnetic interface plays an essential role in molecular spintronics[12], but intense surface activity localizes and even dissociates adsorbed molecules[13]. Experimentally, 2D buffer layers such as graphene[14] and h-BN[15] are applied to passivate and redecorate the ferromagnetic surfaces. The exploration of a 2D robust nanoporous architecture as a decoupling and templating layer for the tailoring of a ferromagnetic surface potential landscape remains uncharted.

In this article, we propose a protocol for self-assembling a 2D-MOPN on an ultrathin ferromagnetic film. On an ultrathin Co film (5 monolayers (ML)), the deposited 2,4,6-tris(4-pyridyl)-1,3,5triazine (T4PT) molecules can coordinate surface Co atoms to form an ordered Co-T4PT porous network. The large-scale porous architecture reshapes the Co surface potential, thus tailoring the arrangement of the subsequently adsorbed $C_{60}$ molecules.

## Results and Discussions

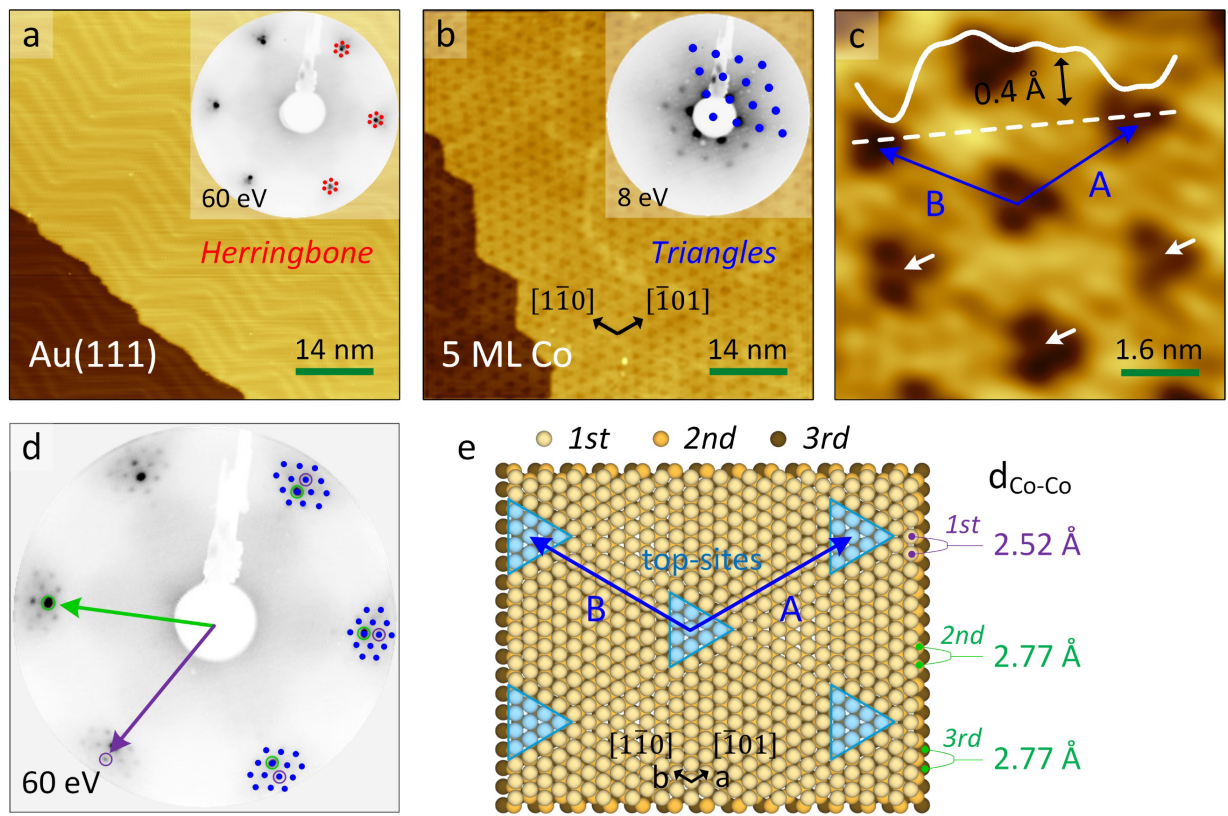

**Figure 1.** (**a**) STM topography of Au(111), showing the surface herringbone reconstruction. The corresponding LEED structure (at 60 eV) is overlapped by the herringbone lattice of $(22 \times \sqrt{3})$ (red dots in the right part). (**b**) STM topography of 5 monolayers (ML) Co on Au(111) shows a triangular surface reconstruction. The LEED image at 8 eV indicates the reconstructed lattice of $(11 \times 11)$, overlapped by the blue dots in the right corner. (**c**) High-resolved STM resolves triangular structures with a lattice of $|\mathbf{A}| = |\mathbf{B}| = 28 \pm 2$ Å and an average depth of 0.4 Å in the line profile. The white arrows indicate clusters in the triangles. (**d**) LEED image of the 5 ML Co at 60 eV shows multiple diffraction spots from the triangular reconstruction (blue dots), first layer (1st) Co lattice (purple circles) and underneath layers Co lattice (green circles). (**e**) The proposed model of the top three layers of the Co. The labeled Co lattice ($d_{Co-Co}$) is 2.52 Å in 1st layer and 2.77 Å in the second (2nd) and third (3rd) layers. The blue triangular regions mark the Co atoms with top-sites stacking in the layers, and they form a periodic pattern with the vectors of **A** and **B** corresponding to the triangular lattice in (**c**). All STM images were recorded at 297 K, and the tip bias ($V_{tip}$) was + 1.02 V in (**a**) and + 0.79 V in (**b, c**).

The cleaned Au(111) surface in Figure 1a shows the well-ordered herringbone reconstruction[16], in which the bright waves are generated by the surface stacking faults of face-centered cubic (FCC) and hexagonal close-packed (HCP). The corresponding LEED image indicates the diffraction structures (red dots) of the reconstructed lattice of the $(22 \times \sqrt{3})$. After further deposition of 5 monolayers (ML) Co on the Au surface, the grown films are annealed at 560 K to create a the large-scale surface terrace. As shown in Figure 1b, the post-annealed Co surface forms the massive triangular reconstructions, and the LEED image at 8 eV exhibits the first-orders diffraction spots of the periodic triangles. The highly resolved STM topography in Figure 1c resolves the reconstructed units with a lattice of $|\mathbf{A}| = |\mathbf{B}| = 28 \pm 2$ Å along the surface <1$\bar{1}$0> directions, and the line profile across the dark triangles shows an average depth of 0.4 Å. These indicate the triangular reconstruction resulting from a lattice mismatch of the surface Co layers. It is well known that the lattice of a Co(0001) crystal (2.50 Å) is very different from that of an Au(111) surface (2.88 Å), and the post-annealed Co films on Au(111) can form a pseudomorphic growth (lattice-matched epitaxy) in the first few layers[17,18]. The annealing temperature (< 575 K) limits the diffusion of the substrate Au atoms in the Co films, but facilities a surface relaxation of the top Co layer.

In Figure 1d, the LEED image at 60 eV shows multiple diffraction structures, where the highest-intensity spots (green circles) are from a lattice of 2.77 Å (close to Au(111) surface lattice) and the peripherally adjacent spots (purple circles) are from a lattice of 2.52 Å (close to Co(0001) crystal lattice). Therefore, a model of the relaxation Co layers is proposed in Figure 1e. The atoms in the top Co layer retain the symmetry (no rotation) but have a smaller lattice of 2.52 Å than the layers below (2.77 Å). The lattice mismatch induces a stacking fault between the layers, and the blue triangular regions indicate the top-sites stacking Co atoms forming a superlattice of $(11 \times 11)\,R0°$. The stacking patterns have a size and a lattice consistent with the triangular reconstruction in Figure 1c, and the reciprocal structures of the $(11 \times 11)\,R0°$ are well superimposed on the diffraction spots (blue dots) in Figure 1b and 1d.

Furthermore, a reciprocal diffraction simulation is performed on the model of the top three Co layers in Figure S1, and the simulated spots reproduce the intensities and positions of the experimental LEED. However, there are some differences between the triangular structures in Figure 1c and the typical moiré patterns[19] because some dark triangles vary in size and exhibit a cluster structure in the centers (indicated by white arrows). These characters indicate that the triangles on the Co surface are from the annealing-induced triangular dislocation loops (TDLs)[20-22]. At the top sites, the stacking atoms in the top layer can squeeze out the underlying atoms because of strain relief, creating some atomic vacancies in the second Co layer. Therefore, the generations of TDLs maintain consistency with the stacking regions at the surface top sites.

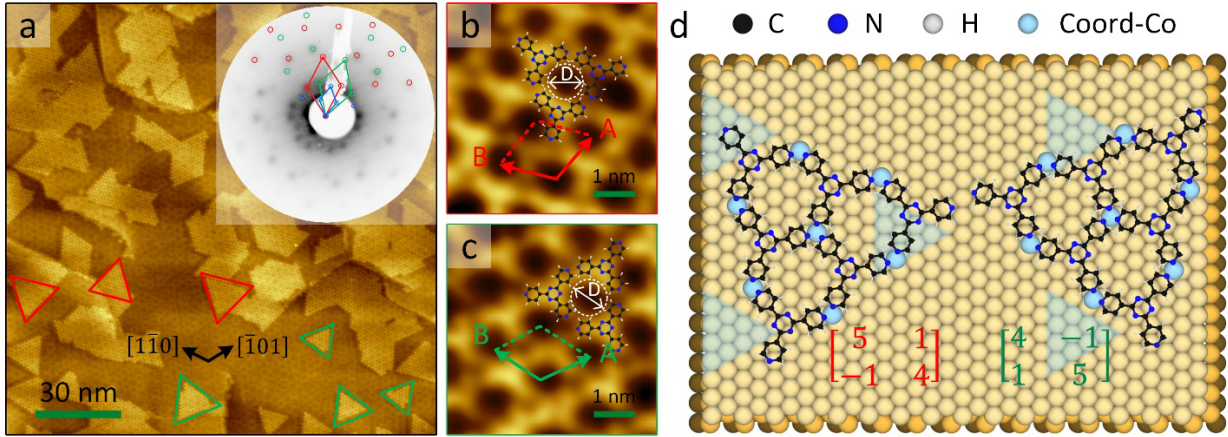

**Figure 2.** (a) Large-scale STM topography of 0.4 ML Co-T4PT network grown on 5 ML Co surface. The two types of Co-T4PT networks (red and green triangles) are marked corresponding to the mirror diffraction structures (red and green rhombi) in the LEED image (at 18 eV). The blue circles are from the underlying TDL lattice. (b, c) High-resolved STM topographies show the two Co-T4PT porous structures. As shown in the superimposed molecular models, each cavity is composed of three T4PT molecules and has a diameter of D = 7.5 ± 0.5 Å. The mirror superlattices are $\begin{bmatrix} 5 & 1 \\ -1 & 4 \end{bmatrix}$ (red unit cell) and $\begin{bmatrix} 4 & -1 \\ 1 & 5 \end{bmatrix}$ (green unit cell), respectively, corresponding to the overlapping circles in the LEED image. (d) The proposed model of the Co-T4PT mirror networks on the Co surface shows the neighboring T4PT molecules coordinated by a Co adatom (Coord-Co). The blue triangles represent the triangular dislocation loops of the Co surface. All STM were acquired at 106 K, and the $V_{tip}$ was + 0.40 V in (a) and + 0.05 V in (b, c).

On the (111)-terminated coinage metal surfaces (Cu and Au), the adsorbed T4PT molecules with the nitrogen ligands spontaneously coordinate with the surface metal adatoms, self-assembling into 2D-MOPNs[23-24]. Here, we deposit 0.4 ML T4PT on the 5 ML Co surface, where the as-grown molecules in Figure S2a show some disordered coordinations. A slight annealing at 413 K generates sufficient kinetic energy for molecules and surface Co adatoms (escaping from the TDLs) to fabricate the Co-T4PT porous networks, as depicted in Figure 2a. The LEED image (red and green circles) indicates the mirror structures on the TDL surface, and the red and green triangles in the STM topography identify the two types of triangular networks.

The small-scale STM topographies in Figure 2b and 2c resolve the Co-T4PT mirror lattices: the porous superstructures are described by the matrices of $\begin{bmatrix} 5 & 1 \\ -1 & 4 \end{bmatrix}$ and $\begin{bmatrix} 4 & -1 \\ 1 & 5 \end{bmatrix}$, and the unit cell parameters are |**A**| = |**B**| = 11.7 ± 0.6 Å along the ±11° to the <1$\bar{1}$0> directions. Each cavity is surrounded by three T4PT molecules and the porous size is D = 7.5 ± 0.5 Å, as indicated in the figures. Figure 2d shows the model of the Co-T4PT mirror networks on the Co surface. The neighboring T4PT molecules are coordinated by a Co atom (coord-Co), which is not visible in the STM images as a fragile feature of the single metal-coordinated center[25]. In the mirror porous networks, each T4PT molecule is adsorbed at the equivalent position of the Co surface. Due to the intact atomic structure in the top Co layer, the surface TDLs have a negligible effect on the Co-T4PT networks.

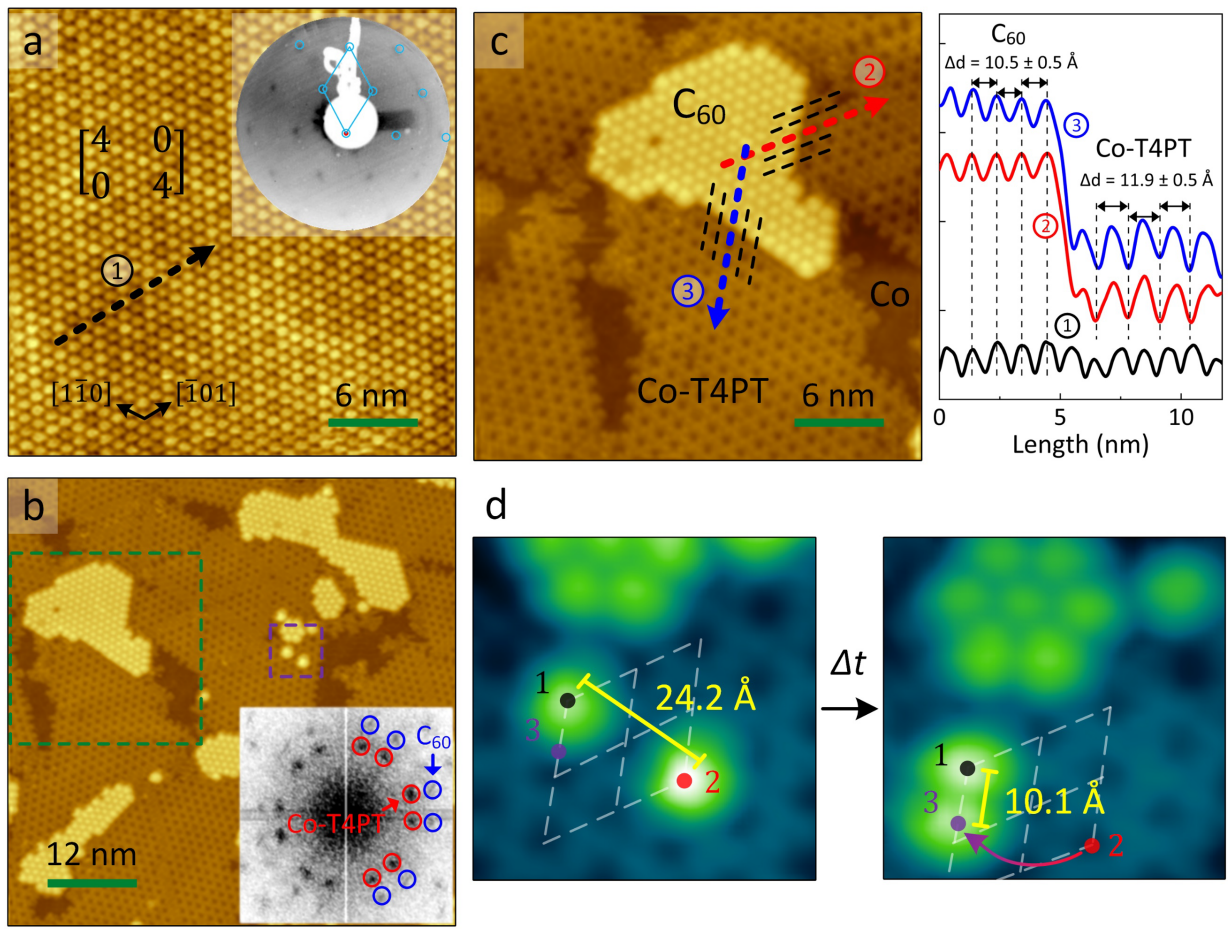

**Figure 3**. (a) STM topography of 1 ML $C_{60}$ on 5 ML Co surface. The corresponding LEED image indicates a (4×4) $R0°$ lattice of the $C_{60}$, overlapped by the light blue circles. (b) STM topography of 0.15 ML $C_{60}$ on 0.9 ML Co-T4PT network on 5 ML Co surface. The image of the lower-right fast Fourier transform (FFT) (18 × 18 $nm^{-2}$) points out two types of structures from the $C_{60}$ (out spots marked by the blue circles) and the Co-T4PT (inner spots marked by the red circles) lattices. (c) The zoom-in STM image of the green dashed rectangle in (b). The short-dashed lines indicate the adsorbed $C_{60}$ molecules following the porous symmetry of the underlying Co-T4PT. In the right part, the line profiles along the color arrows in the structures of (a) and (c). The lattice distances (Δd) of the $C_{60}$ and the Co-T4PT are 10.5 ± 0.5 Å and 11.9 ± 0.5 Å, respectively. (d) Zoom of the purple dashed rectangle in (b). The left and right images are the two real-time STM recorded at a frame interval Δt = 419 s. Left image, two $C_{60}$ molecules separate at sites "1" and "2" with a distance of 24.2 Å. On the Right image, the two molecules aggregate at the sites "1" and "3" with a distance of 10.1 Å. The white grids show the Co-T4PT porous positions. All STM images were recorded at 106 K, and the $V_{tip}$ was + 0.22 V in (a) and + 0.54 V in (b)-(d).

In a 2D-MOPN, the metal-coordinated organic molecules can create potential barriers to block the penetration of surface electrons, adatoms and even molecules[5,26]. The well-ordered organic porous architectures form a periodic array of quantum wells through the organic building blocks to trap further adsorbates. For instances, the 2D porous architectures of the BDBA network[27] and the NN4A network[28] are fabricated on the HOPG surface. Each constructed cavity has a porous size of ~ 8 Å, and the further deposited $C_{60}$ molecules are trapped in the cavities forming an ordered $C_{60}$ array. Due to the porous confinement effect, large-scale 2D-MOPN can reshape the entire surface potential landscape. In the following, we explore the behaviors of $C_{60}$ molecules in the Co-T4PT network on the Co surface.

To understand the surface potential landscape in 5 ML Co, 1 ML $C_{60}$ molecules are initially deposited on the Co surface, as shown in Figure 3a. The STM topography and the LEED image indicate a commensurate superstructure described by the $(4 \times 4)$ $R0°$, and the lattice is $10.5 \pm 0.5$ Å along the surface <1$\bar{1}$0> directions. In a $C_{60}$ film, the strong interaction between the molecules[29] results in an intermolecular distance equivalent to the $C_{60}$ van der Waals (vdW) diameter (~ 10 Å)[30]. Therefore, the $C_{60}$ molecules adopt a $(4 \times 4)$ $R0°$ configuration on graphene and Cu(111) surfaces[31,32] with a substrate lattice of ~ 2.5 Å, while they form a $(2\sqrt{3} \times 2\sqrt{3})$ $R30°$ configuration on Au(111) and Ag(111) surfaces[33] with a substrate lattice of ~ 2.8 Å. In contrast, the $(4 \times 4)$ superlattice on the Co surface reflects that the atomic distance ($d_{Co-Co}$) in the top Co layer is about 2.5 Å, in agreement with Figure 1e. Furthermore, the STM topography in Figure S3 demonstrates the superlattice matching between the Co surface TDLs $(11 \times 11)$ and the adsorbed $C_{60}$ molecules $(4 \times 4)$.

In Figure 3b, the Co surface is first covered with a 0.9 ML Co-T4PT porous network, followed by the deposition of 0.15 ML $C_{60}$ molecules. The $C_{60}$ molecules tend to aggregate into ordered arrays at the boundaries of the Co-T4PT networks. The corresponding fast Fourier transform (FFT) image reveals two sets of spots with identical symmetry: the inner ones (red circles) originate from the Co-T4PT networks, while the outer ones (blue circles) stem from the $C_{60}$ arrays. This indicates the different lattices between the two structures. The green dashed region is magnified in Figure 3c. The short-dashed lines depict the $C_{60}$ array following the lattice directions of the underlying network, but the molecules are not precisely located in the Co-T4PT cavities. The line profiles along the red and blue arrows show a smaller lattice of the $C_{60}$ array ($10.5 \pm 0.5$ Å) than the underlying porous network ($11.9 \pm 0.5$ Å). Remarkably, the lattice of the $C_{60}$ array on the Co-T4PT network is identical to that on the Co surface (line profile along the black arrow), indicating a significant interaction between the $C_{60}$ molecules.

Figure 3d shows two real-time STM images recorded in the purple dashed region in Figure 3b. In the left image, two isolated $C_{60}$ molecules are located in the Co-T4PT cavities of "1" and "2" with a distance of 24.2 Å, almost twice the interporous distance (~ 12 Å). After a frame interval Δt = 419 s, a $C_{60}$ molecule moves from position "2" to form a dimer with another molecule at position "3", as depicted in the image on the right. The distance in the dimer measures 10.1 Å, which is in line with the vdW diameter of $C_{60}$ but smaller than the interporous distance. As a result, the interacting $C_{60}$ molecules are out of the cavities. The different cases occur in the mentioned BDBA and NN4A networks, where the interporous distances are 15 Å and 26 Å, respectively, beyond the $C_{60}$ interacting diameter. The porous potential dominates the adsorbed $C_{60}$ molecules precisely in their cavities. The Co-T4PT network has a suitable cavity size (D = 7.5 ± 0.5 Å) for the molecular confinement, but the shorter cavity distance (11.9 ± 0.5 Å) is in the range of the $C_{60}$ interacting diameter. Therefore, the arrangement of the $C_{60}$ molecules is steered by the competition between the Co-T4PT porous potential and the vdW interactions of the neighboring $C_{60}$.

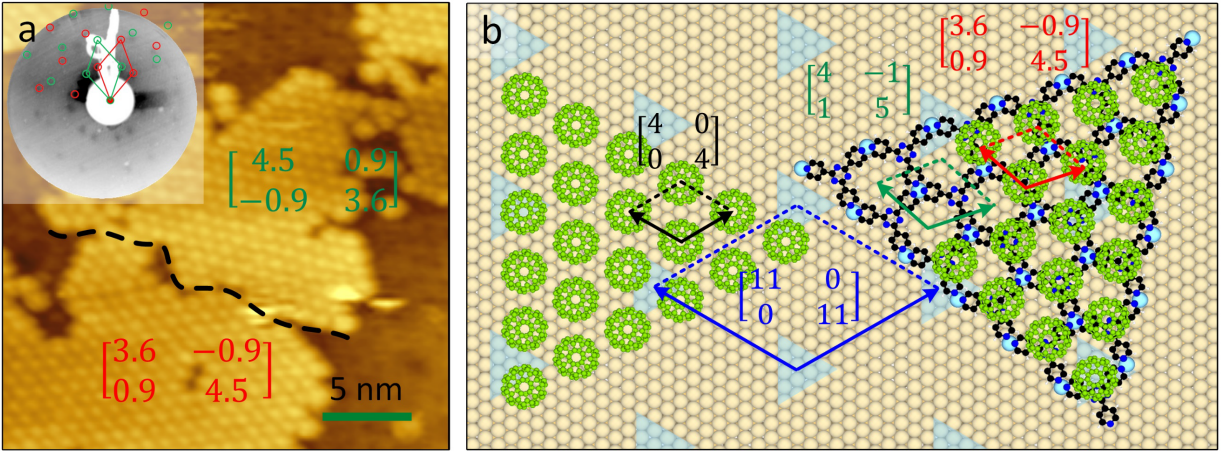

**Figure 4**. (**a**) STM topography of 0.75 ML $C_{60}$ on 0.9 ML Co-T4PT network on 5 ML Co surface. The $C_{60}$ domains separated by the black dashed line show the mirror structures of $\begin{bmatrix} 4.5 & 0.9 \\ -0.9 & 3.6 \end{bmatrix}$ and $\begin{bmatrix} 3.6 & -0.9 \\ 0.9 & 4.5 \end{bmatrix}$. The simulated diffraction spots (green and red circles) are superimposed on the corresponding LEED image (at 18 eV). (**b**) The proposed model shows the structures of $C_{60}$ molecules adsorbed on the Co surface and on the Co-T4PT network, and the corresponding unit cells and lattice matrices are marked. The blue triangles represent the triangular dislocation loops of the Co surface. The STM was recorded at 106 K and the $V_{tip}$ was + 0.50V in (**a**).

When 0.75 ML $C_{60}$ is covered on the Co-T4PT network in Figure 4a, the molecules conform to the symmetries of the underlying networks and aggregate into the large domains. The corresponding LEED image (upper-left corner) also indicates that the $C_{60}$ configurations have the same shape as the Co-T4PT networks (in Figure 2a), but form the smaller superlattices described by the matrices of $\begin{bmatrix} 4.5 & 0.9 \\ -0.9 & 3.6 \end{bmatrix}$ and $\begin{bmatrix} 3.6 & -0.9 \\ 0.9 & 4.5 \end{bmatrix}$. The structures of $C_{60}$ molecules adsorbed on the Co surface and on the Co-T4PT network are illustrated in Figure 4b, with the lattice parameters summarized in Table S1. By comparing the adsorption of $C_{60}$ on both surfaces, the two-dimensional Co-T4PT network has been substantiated as a templating layer to tailor the ferromagnetic Co surface potential landscape.

**Conclusions**

The ultrathin Co films are grown on the Au(111) surface, and the lattice mismatch of the top two layers results in a well-ordered TDL reconstruction. Supported on the Co surface, the deposited T4PT molecules can coordinate with the surface Co atoms to form a large-scale Co-T4PT porous network. The Co-T4PT network contains periodic nanocavities as a quantum well array to reshape the surface potential landscape of ferromagnetic Co. Subsequently adsorbed $C_{60}$ molecules arrange along the symmetry of the underlying porous networks, but their interactions with neighbouring molecules cause them to deviate from the cavities, resulting in a smaller lattice. Therefore, the $C_{60}$ configurations are dominated by the Co-T4PT porous potential and the neighbouring $C_{60}$ interactions. Our research proposes a functionalized protocol for 2D-MOPNs on ferromagnetic surfaces that can be used as templates for molecule-based electronics and spintronics.

**Methods**

**Sample preparation.** All experiments were performed in ultra-high-vacuum systems (base pressure $10^{-10}$ mbar) housed on an Omicron MULTIPROBE STM. Prior to deposition of adsorbates, the Au(111) surface was cleaned by several cycles of Ar-ion sputtering and subsequent annealing. The surface cleanliness was inspected using the LEED and the STM images, as shown in Figure 1a. Co evaporation was conducted by an electron beam evaporator (FOCUS EFM3) with a flux of 40 nA, and the coverage was determined by calculating a time-integrated thickness of the as-grown submonolayer in STM. The grown Co films were annealed at 560 K for 15 mins to form large-scale surface terraces[17]. The purified T4PT and $C_{60}$ molecules were thermally evaporated at 450 K and 638 K, respectively, while the samples were at room temperature. The molecular deposition rates were 0.25 ML/min for T4PT and 0.06 ML/min for $C_{60}$ monitored by a quartz crystal microbalance (QCM), and the coverage was calibrated by the STM. The T4PT molecules deposited on the 5 ML Co were annealed at 413 K for 60 mins to fabricate the Co-T4PT network. The grown $C_{60}$ films were not further annealed in the experiments.

**Scanning tunneling microscopy (STM).** The STM setup has been reported in previous work[34]. In brief, all STM topographies were recorded in constant current mode with a tunneling current ($I_t$) in the range of 70-90 pA. The bias voltage ($V_{tip}$) was applied to the STM tip, where positive and negative values corresponded to the occupied and unoccupied states of the samples, respectively. The low-temperature STM experiments at 106 K were conducted using a liquid $N_2$ system. The STM data were processed using the WSxM and the Gwyddion softwares[35,36].

**Low energy electron diffraction (LEED).** The LEED images were obtained using a rear view four grid system at room temperature, and some diffraction spots were overlaid with the corresponding matrices simulated in Spot-Plotter Software[37].

**Acknowledgments**

This work was supported by the Deutsche Forschungsgemeinschaft (DFG, German Research Foundation), TRR 173-268565370 Spin + X: spin in its collective environment (Project B05). B.S. acknowledges financial support by the Dynamics and Topology Center funded by the State of Rhineland Palatinate.

# Tailoring the ferromagnetic surface potential landscape by a templating two-dimensional metal-organic network

by
L. Lyu et al.

**Supplementary Figures**

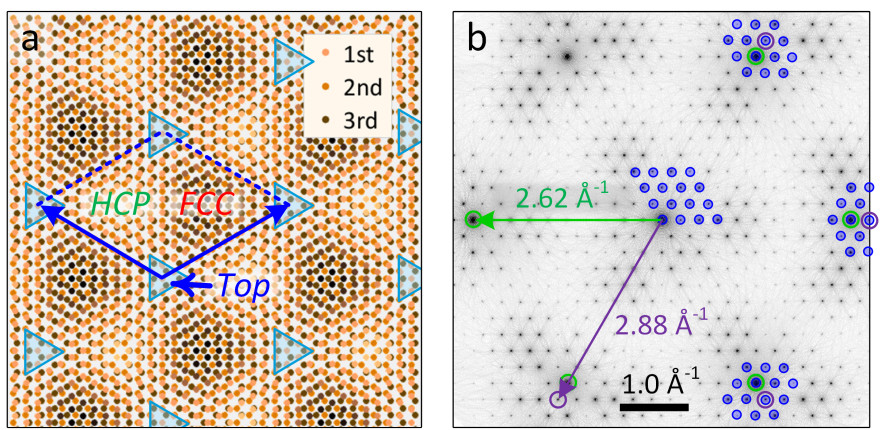

**Figure S1**. (**a**) The model of the top three layers Co for the LEED simulation. The Co atoms have the same symmetry in each layer, but the $d_{Co-Co}$ lattices are 2.52 Å in the 1st layer and 2.77 Å in the 2nd and 3rd layers. The three types of atom stacking faults are found in the surface regions: HCP (A-B-A), FCC (A-B-C) and Top (A-A-A). The blue triangles mark the top stacking regions, forming a (11 × 11) lattice (the blue rhombus). (**b**) The simulated diffraction structures, including the reciprocal lattices of the 1st layer Co (purple circles, $\frac{2\pi}{2.52} \times \frac{2}{\sqrt{3}}$ = 2.88 Å$^{-1}$), the 2nd layer Co (green circles, $\frac{2\pi}{2.77} \times \frac{2}{\sqrt{3}}$ = 2.62 Å$^{-1}$) and the atom stacking faults (blue circles, 11 × 11). The diffraction simulation was performed in the Matlab program[S1].

[S1] Martoccia, D. et al. Graphene on Ru(0001): a corrugated and chiral structure. *New Journal of Physics* **12**, 043028 (2010).

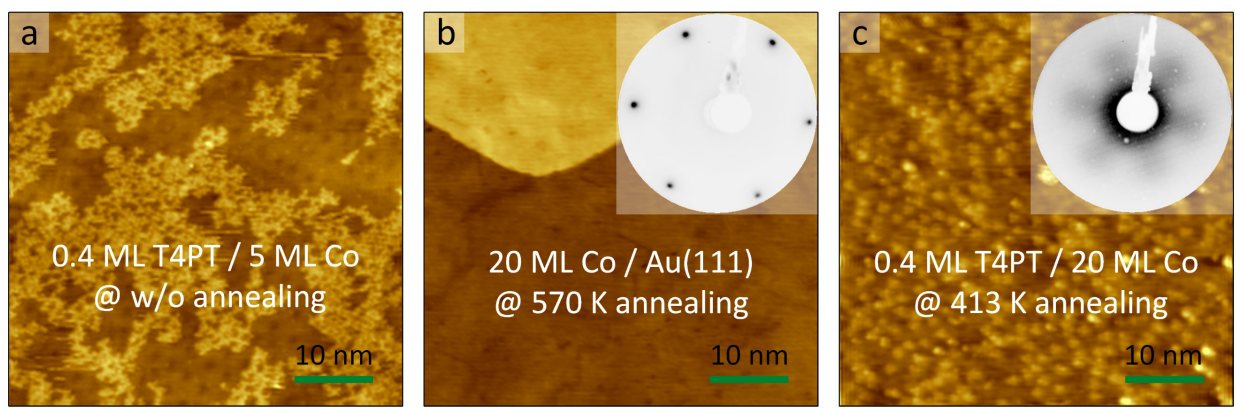

**Figure S2**. (**a**) STM topography of 0.4 ML T4PT on 5 ML Co surface prior to the annealing process (w/o annealing). (**b**) STM topography of 20 ML Co on Au(111) after annealing of 560 K, and the upper-left LEED image (at 60 eV) indicates the bulk-phase Co structure. (**c**) STM topography of 0.4 ML T4PT on 20 ML Co surface after annealing of 413 K. The T4PT molecules are disordered on the surface, and the corresponding LEED image (at 12 eV) shows the blurred spots. All STM images were recorded at 106 K, and the $V_{tip}$ was + 0.71 V in (**a**), + 0.29 V in (**b**) and + 0.95 V in (**c**).

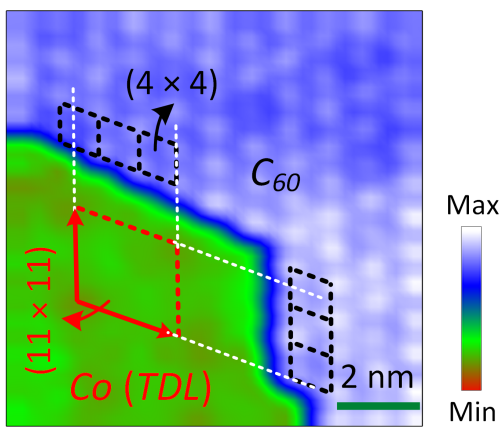

**Figure S3**. STM image of 0.6 ML $C_{60}$ on 5 ML Co surface. The red rhombus is the unit cell of the Co TDL structure (11 × 11), which extends the lattice (white dashed lines) to the adsorbed $C_{60}$ unit cells (black rhombi, 4 × 4). The STM was acquired at 106 K, and the $V_{tip}$ was – 0.08 V.

**Table S1**. Summary of the structural parameters (unit cell and lattice matrix) for 5 ML Co / Au(111), $C_{60}$ / 5 ML Co, Co-T4PT / 5 ML Co, and $C_{60}$ / Co-T4PT / 5 ML Co. The lattice matrices are obtained by fitting the LEED spots. Each unit cell is calibrated by the corresponding matrix, excluding reasonable errors. γ is the angle of the lattice vectors <**a**,**b**> and <**A**,**B**>. The basis vectors for all the matrices are |**a**| = |**b**| = 2.52 Å and γ = 120°.

| Geometry | Unit Cell | Matrix |
|---|---|---|
| 5 ML Co / Au(111) | \|**a**\| = \|**b**\| = 2.52 Å, γ = 120° | $\begin{bmatrix} 11 & 0 \\ 0 & 11 \end{bmatrix}$ (TDL) |
| | \|**A**\| = \|**B**\| = 27.72 Å, γ = 120° (TDL) | |
| $C_{60}$ / 5 ML Co / Au(111) | \|**A**\| = \|**B**\| = 10.08 Å, γ = 120° | $\begin{bmatrix} 4 & 0 \\ 0 & 4 \end{bmatrix}$ |
| Co-T4PT / 5 ML Co / Au(111) | \|**A**\| = \|**B**\| = 11.55 Å, γ = 120° | $\begin{bmatrix} 5 & 1 \\ -1 & 4 \end{bmatrix} / \begin{bmatrix} 4 & -1 \\ 1 & 5 \end{bmatrix}$ |
| $C_{60}$ / Co-T4PT / 5 ML Co / Au(111) | \|**A**\| = \|**B**\| = 10.39 Å, γ = 120° | $\begin{bmatrix} 4.5 & 0.9 \\ -0.9 & 3.6 \end{bmatrix} / \begin{bmatrix} 3.6 & -0.9 \\ 0.9 & 4.5 \end{bmatrix}$ |